# Low-defect quantum dot lasers directly grown on silicon exhibiting low threshold current and high output power at elevated temperatures.

*Konstantinos Papatryfonos[1,2]\*, Jean-Christophe Girard[2]\*, Mingchu Tang[1], Huiwen Deng[1], Alwyn J. Seeds[1], Christophe David[2], Guillemin Rodary[2], Huiyun Liu[1], David R. Selviah[1]*

Abstract

The direct growth of III-V materials on silicon is a key enabler for developing monolithically integrated lasers, offering substantial potential for ultra-dense photonic integration in vital communications and computing technologies. However, the III-V/Si lattice and thermal expansion mismatch pose significant hurdles, leading to defects that degrade lasing performance. This study overcomes this challenge, demonstrating InAs/GaAs-on-Si lasers that perform on par with top-tier lasers on native GaAs substrates. This is achieved through a newly developed epitaxial approach comprising a series of rigorously optimised growth strategies. Atomic-resolution scanning tunnelling microscopy and spectroscopy experiments reveal exceptional material quality in the active region, and elucidate the impact of each growth strategy on defect dynamics. The optimised III-V-on-silicon ridge-waveguide lasers demonstrate a continuous-wave threshold current as low as 6 mA and high-temperature operation reaching 165 °C. At 80 °C, critical for data centre applications, they maintain a 12-mA threshold and 35 mW output power. Furthermore, lasers fabricated on both Si and GaAs substrates using identical processes exhibit virtually identical average threshold current. By eliminating the performance limitations associated with the GaAs/Si mismatch, this study paves the way for robust and high-density integration of a broad spectrum of critical III-V photonic technologies into the silicon ecosystem.

## 1. Introduction

Silicon electronics, based on complementary metal-oxide-semiconductor (CMOS) technology, has revolutionized modern life by enabling the ubiquitous use of silicon chips in computers, phones, and data centres (DCs). In parallel, silicon photonics (SiP) has emerged with the ambitious goal of leveraging this vast silicon infrastructure by integrating it with other materials that can unlock new functionalities, such as III-V materials for photonic integrated circuits.[1,2,3] Beyond its inherent compatibility with CMOS processes, SiP technology offers a plethora of advantages including low power consumption, compact building blocks, and low crosstalk, positioning it as a promising platform for manipulating and controlling light signals near or on CMOS circuitry.[4,5,6,7,8,9] The integration of photonic and electronic circuits on a single silicon chip holds immense potential for significantly reducing system complexity, size, and cost while maintaining the advantages of both electronic and photonic signal processing. This presents a compelling avenue to address critical technological challenges across various domains, including data communications (datacom), telecommunications, computing, and sensing.[4,10]

At the heart of SiP's development lies the core objective of integrating highly efficient and reliable light sources on silicon.[7,8,11] In particular, high-performance integrated lasers are a long sought-after goal, expected to play an important role in the development of efficient optical interconnects for data centres, crucial for handling the ever-increasing demands of cloud applications, social media, and big data analytics.[7,9] To this end, a significant milestone is achieving Si-based lasers that can maintain high performance levels even at elevated temperatures (≥ 80 °C), as expected in data centre input/output (I/O) chips.[9,11] The realization of this advancement would mark a crucial step towards replacing external light sources and moving towards fully-integrated architectures.[8,9] Silicon, with its indirect bandgap, is not an efficient light-emitting material.[12,13,14] To surmount this limitation, the integration of III-V materials, renowned for their excellent light-emitting properties, with silicon has emerged as the most promising pathway.[8,13,15] The two prominent methods under exploration are direct growth[16,17,18,19] and wafer bonding[20,21].

---
[1] Department of Electronic and Electrical Engineering, University College London, London WC1E 7JE, UK.
[2] Université Paris-Saclay, CNRS, Centre de Nanosciences et de Nanotechnologies (C2N), 91120 Palaiseau, France.
\*email: k.papatryfonos@ucl.ac.uk , konstantinos.papatryfonos@c2n.upsaclay.fr ; jean-christophe.girard@c2n.upsaclay.fr

While III-V-on-silicon (III-V-on-Si) wafer bonding techniques have been extensively researched and shown to deliver satisfactory performance,[4,7,22,23,24] the direct III-V growth on Si may offer compelling advantages if high performance can be maintained.[8,25] Notably, it promises cost-effectiveness,[11] scalability for mass production,[26] and a reduced environmental impact due to the elimination of III-V substrate waste. However, the large lattice mismatch and thermal expansion mismatch between III-Vs and Si introduces strain-induced structural defects, such as threading (TD) and misfit (MD) dislocations, which can significantly affect device performance.[15,27] Minimizing the defect density together with utilizing optimised quantum dots[28,29] (QDs) as the active material, which can enable a higher temperature insensitivity of the threshold current,[28,30] are key strategies for improving performance at and above 80 °C. QD-based III-V-on-Si lasers were recently shown to also exhibit a high reliability, in terms of a long mean time to failure, at both room temperature[8] and 80 °C[11].

In this study, we demonstrate the direct growth and fabrication of III-V-on-Si quantum dot lasers that exhibit a low threshold current and high output power at elevated operating temperatures. The enhanced material quality of our devices stems from a newly developed epitaxial approach that combines a series of highly refined molecular beam epitaxy (MBE) growth strategies, resulting in an exceptionally low defect density in the active region. To rigorously assess our epitaxial methodology, we employ atomic-resolution cross-sectional scanning tunnelling microscopy and spectroscopy (X-STM/STS). This cutting-edge technique provides invaluable insights into the material quality and the electronic structure of the quantum dot-in-a-well (DWELL) active region. It also enables real-space squared wave function imaging and detailed tracking of defect evolution, crucial for understanding the influence of our growth strategies. Our findings indicate a defect density below $10^5$ cm$^{-2}$ in the active region, approaching the theoretical limit.[31,32] This represents a significant advancement of III-V/Si heteroepitaxy, particularly considering the large lattice mismatch between silicon substrates and III-V compound semiconductor buffer layers.

To evaluate the III-V-on-Si device performance, we fabricate both broad area and narrow ridge-waveguide lasers operating at λ = 1.3 μm. The broad area lasers exhibit an exceptionally low continuous-wave (CW) threshold current density ($J_{th}$) of 48 A cm$^{-2}$, which compares favourably with many recent pioneering reports.[8,25,33,34,35] The narrow-ridge lasers, optimised for single-transverse-mode operation, achieve a CW threshold current as low as 6 mA at room temperature (RT) and 12 mA at 80 °C, while also reaching a high maximum output power of 35 mW at 80 °C. Notably, these lasers maintain ground state lasing up to a maximum temperature of 165 °C, improving on previously reported III-V-on-Si records by more than 40 degrees[8,11]. Furthermore, a direct comparison between our III-V-on-Si and III-V-on-III-V lasers fabricated in parallel using the same process, reveals comparable performances including a virtually identical average threshold current. This marks a significant milestone in III-V monolithic integration on Si with potential to unlock a wide range of applications.

## 2. Results

### 2.1 Cross-sectional Scanning Tunnelling Microscopy and Spectroscopy after *in situ* cleavage of the III-V-on-Si laser.

The MBE-grown epitaxial structure of our III-V-on-Si laser is schematically illustrated in Fig. 1a,b. Detailed descriptions of the growth process and optimization strategies are provided in the Methods section. Our initial analysis involved a comprehensive layer-by-layer examination of the entire structure using X-STM/STS, a technique that uniquely allows simultaneous probing of the physical and electronic structure with atomic resolution for the former and high energy resolution (~30 meV at 77K) for the latter (see Methods for further details). We systematically identified and scanned each interface within the structure using controlled step displacements of the STM tip, following the nominal interspace distance between the layers as prescribed by the MBE growth. Figure 1 and Figures S1-S3 in the Supplementary Information (SI) highlight key examples from this analysis. This approach allowed us to gain detailed insights into how each layer affects dislocation behaviours during growth, which is invaluable for optimising the epitaxial structure. Particular attention was given to the X-STM/STS analysis of the strained-layer superlattices (SLS), known for their efficacy in reducing the defect density propagating towards the active region as successive layers are grown.[8,11,31,32] Multiple InGaAs/GaAs SLS were employed as dislocation filter layers (DFLs) in our structure, while additional AlGaAs/GaAs superlattices were grown above the DFLs as well as below and above the active region to further enhance material quality as depicted in Fig. 1b and elaborated in Methods. We meticulously scanned and imaged several areas in each of these superlattices as well as the regions right below and above them.

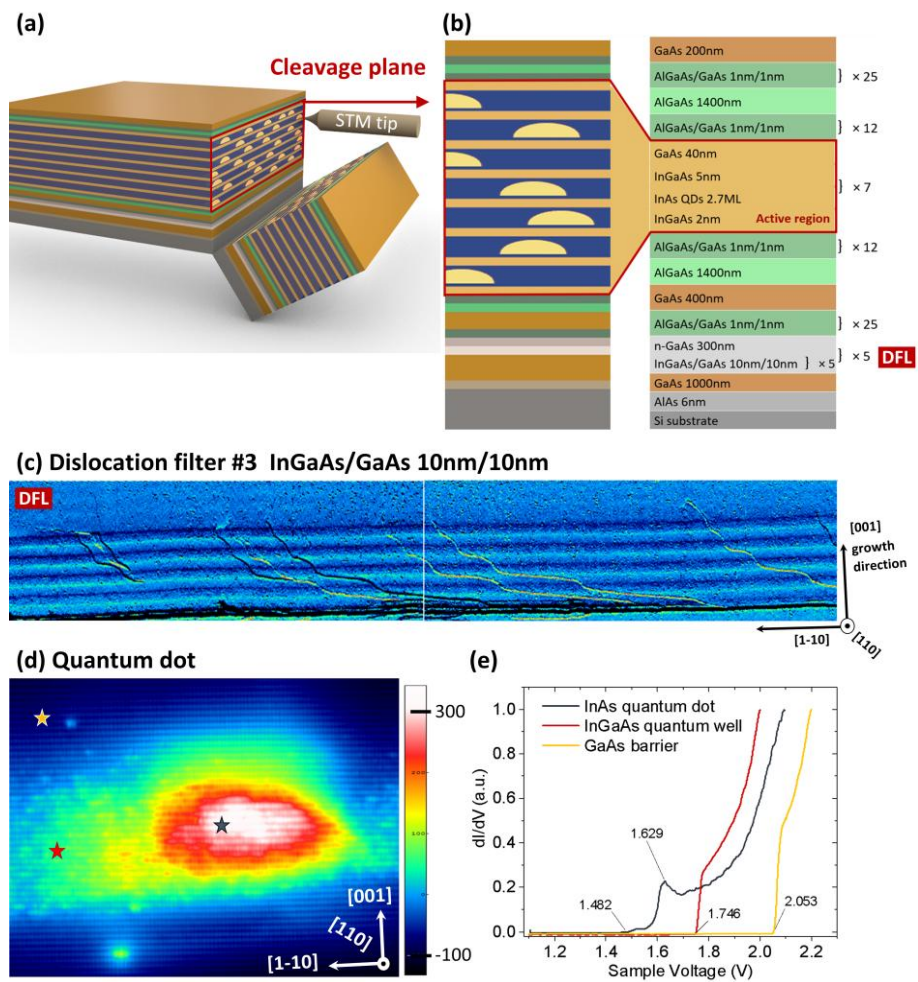

**Figure 1:** (a) Schematic of the cleavage plane of the sample studied by X-STM/STS. (b) Schematic of the epitaxial structure of the optimised III-V-on-Si sample. (c) STM current image of the SLS #3 area (image size: 1000 nm x 170 nm; voltage: -3 V; current: 100 pA). (d) STM topographic image of an individual InAs QD in the stack layer #6, embedded in an In$_{0.16}$Ga$_{0.84}$As QWell and GaAs barrier (27 nm x 38 nm; V = 3 V; I = 40 pA; T = 77K). (e) Individual $\frac{dI}{dV}(V)$ spectra showing the spatial dependence of the onset of the conduction band at the positions denoted in panel (d) with the respectively coloured star symbols. A first shoulder is detected at 1.48 eV, followed by a resonance at 1.62 eV, ascribed to the ground state and the first excited state of the QD. DFL—dislocation filter layers.

Figure 1c presents a typical current image of SLS #3, with the successive horizontal stripes representing the InGaAs/GaAs layers, each measured to be 10 nm thick. The SLS exerts an in-plane force on TDs, thereby enhancing their lateral motion during growth and increasing the likelihood of TD annihilation or deflection. This mechanism effectively traps numerous TDs in layers designed for this purpose, preventing them from reaching the active region. This effect is evident in Figure 1c, where several threading dislocations (visible as black and white curved lines) can be observed propagating through an SLS. As successive strained layers are grown, the dislocations bend and many of them terminate at the end of the superlattice. The defect density estimation above and below the DFLs was conducted by analysing 25 large-scale STM images. To enhance accuracy, we measured the change in defect density after passing through all dislocation filtering layers. Leveraging the well-known initial defect density just above the Si/III-V interface in our samples, this approach allows for a more accurate estimation of the final defect density, which averaged below $10^5$ cm$^{-2}$ in these images. We note that this defect density is lower but remains of the same order as previous TEM analyses performed on similar structures grown in recent years[8,32].

In Figure 1d, a filled-states STM image displays a strong contrast of an individual QD from layer #6, revealing the rows of its arsenic atoms along the [110] crystal direction. From their lattice parameter, we deduced the typical height (growth direction) and width of the QDs, varying between 6-7 nm and 17-22 nm respectively. The surrounding quantum well (QWell) in Fig. 1d is also well-identified in the adjacent green and light-blue regions, though it is more clearly observable in larger-scale images, such as Fig. 2a. We note that dislocations were extremely rare in the QD regions, and no atomic-scale defects or voids[36,37] were observed, signifying their high material quality suitable for lasers and other optoelectronic devices. Figure 1e shows individual differential tunnelling conductance (*dI/dV*) spectra obtained at the positions indicated by the star symbols in Fig. 1d, with the sample voltage *V* varying from 1 V to 2.2 V. Such STS spectra can be used to characterise the electronic confinement in the QDs. The conduction band (CB) threshold is 1.48 V on the QD, 1.75 V on the QWell and 2.05 V on the GaAs barrier, signifying the tunnelling current onset as the energy of the STM tip electrons reaches the lowest available in-gap states before reaching the CB.

The QD spectra exhibit a first shoulder at 1.48 V and then a sharp peak at 1.63 V, corresponding to the ground state and the first excited state resonance, respectively. The excited state peak is more prominent due to the position at which this measurement was taken on the QD (see Fig. 1d). The peaks occur as the tunnelling current abruptly increases when the Fermi level of the tip crosses a discrete QD level, and their amplitudes depend on the local density of states (DOS) at the measurement position.[38,39,40] By taking such *dI/dV* spectra over a dense grid of points we can map the density of states at each position, and so reconstruct the squared wave function in a QD or any selected area, and at any selected energy.[38,39,40] In the following sections, we will discuss such spatially and energetically resolved *dI/dV* maps along with the topographic (structural) properties across a wide area (Fig. 2) and on an individual QD (Fig. 3) in the III-V-on-Si active region.

## 2.2 Simultaneous high-resolution assessment of topographic and electronic properties of the laser active region.

Figure 2a shows a representative topographic image illustrating three quantum dots located in the last two layers (#6 and #7). These QDs are embedded within InGaAs QWells and separated by 37.5 nm GaAs barriers. As depicted in the figure, even these QDs that lie at the higher part of the structure exhibit optimal material quality as no TDs, MDs, cracks, or voids[36,37] are observed in the QDs vicinity whatsoever. These uppermost QD layers are the most critical layers to examine, as they incorporate larger amounts of strain, including that of the QD and QWell layers lying beneath them in the laser stack. It is well-known that such defects produce a very large amount of non-radiative recombination centres and are, thus, detrimental to device performance if they are present at the position where light is produced in the structure. This is, therefore, an important improvement compared to the only other available X-STM topographic study of a similar III-V structure grown on silicon, in which such cracks and voids were indeed observed in the QDs vicinity[36].

Relative to the topographic image shown in Figure 2a, Figures 2b-f depict the wave function mapping over the same large-scale region. Technical details on how we performed the wave function mapping are given in

Methods. In Figure 2b, at V = 1.432 V, we see a single-lobe structure in the QDs, corresponding to the CB ground state. When increasing the energy of the tunnelling electrons, we progressively observe two lobes in Fig. 2c and then three lobes in Fig. 2d, at V = 1.522 V and 1.638 V respectively. There are only three confined states in these QDs, and increasing the energy further results in the wave function progressively becoming less localised and ultimately being extended in the QWell area (Fig. 2e,f), as more states are becoming available within the QWell. Figure 2e, taken at V = 1.734 V, presents a state in which the wave function extends into the wetting layer (WL) and a little bit into the QWell but only in an area around the QD vicinity, whereas Figure 2f shows an even more delocalised state at V = 2.087 V. Most of the quantum dots that we investigated individually, showed three well-confined quantised electronic states with a few exceptions of smaller dots showing two states. Further details on our DWELL wave function profiles are given in the Supplementary Note B.

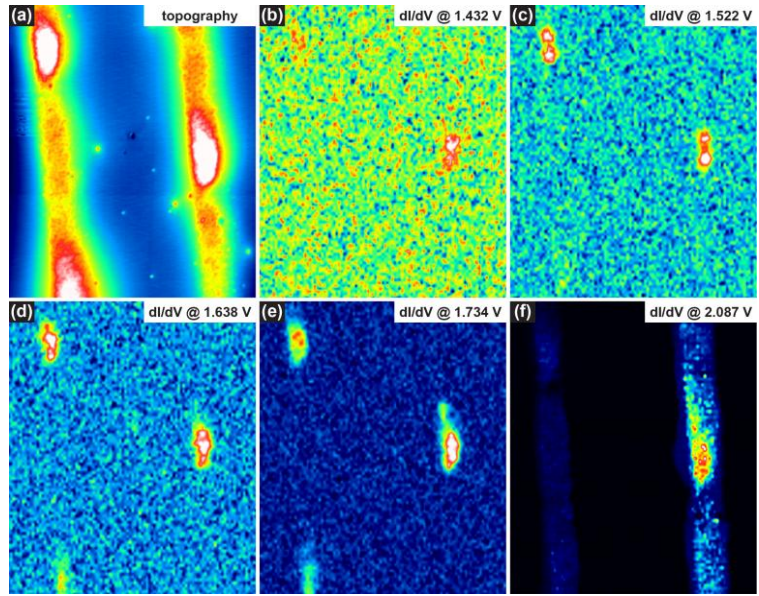

**Figure 2:** (a) Topographic image of QDs in layers #6 and #7 (100 nm x 100 nm; V = -3 V; I = 50 pA); The Z colour scale varies from 0 to 455 pm. (b) – (f) Differential tunnelling conductance (*dI/dV*) maps at sample voltages: V = 1.432 V; V = 1.522 V; V = 1.638 V; V = 1.734 V and V = 2.087 V. (b-d) The QDs exhibit three well-confined states: (b) ground state wave function with a single lobe; (c) first excited state with two lobes and one node at the centre of the QD; (d) second excited state with three lobes and two nodes. (e, f) The wave function is delocalised beyond the QD, extending into (e) the wetting layer, and (f) into the surrounding quantum well.

### 2.3 Squared wave function imaging of an individual III-V-on-Si quantum dot-in-a-well, and finite-element simulations.

Following the same methodology as the previous section, but now focussing on individual QDs, we mapped their electronic wave functions at high-resolution. A representative result for an average-sized QD in layer #7 is illustrated in Figure 3. To support these experimental results, we also performed detailed finite-element simulations, which are generally known to fit well with STM measurements,[39,40,41] as the latter provide the geometry of the structure with atomic resolution which substantially improves the simulation accuracy. Furthermore, STS gives information for the corresponding electronic energy levels structure which can be directly compared to the modelled results. The technical details for implementing our model are provided in Methods. Figure 3 presents typical simulation results alongside their experimental counterparts, with the simulations carried out for a QD of average size cleaved at its centre (in analogy to an STM cleavage), leading to a sub-surface depth of 9 nm (height = 6.5 nm, width = 18 nm, depth = 9 nm). The surrounding quantum well height is set to 9 nm in the simulations, slightly higher than the nominal MBE growth thickness, to ensure that the QDs are always capped by it, in accordance to what we see in the STM analysis. The QD is then placed asymmetrically within the QWell, similarly to the epitaxial structure of Fig. 1b.

As shown in Figures 3a-c, below 1.8 eV, the experimental and simulated wave function maps exhibit the same sequence, corresponding to the (*n00*) bound states (with *n* = 0, 1, 2), where *n* corresponds to the number of nodes in the lateral QD direction, [110]. Above 1.8 eV (Figs. 3d and e), the simulated maps also have a similar structure to the experimental wave function profiles when considering the superposition of adjacent energy levels that are overlapping due to the energy level broadening at 77K. We must note here that when plotting the simulated superposed wave function of multiple (>2) adjacent energy levels, as the ones of Figure 3e, we did not assume an equal weighting, because the contribution of each state depends on its precise energetic distance from the plotted energy (see Methods). The calculated energy levels agree quantitatively with the experimental ones after applying the well-known corrections for the tip-induced band bending (TIBB) effect[72,73] to the latter.

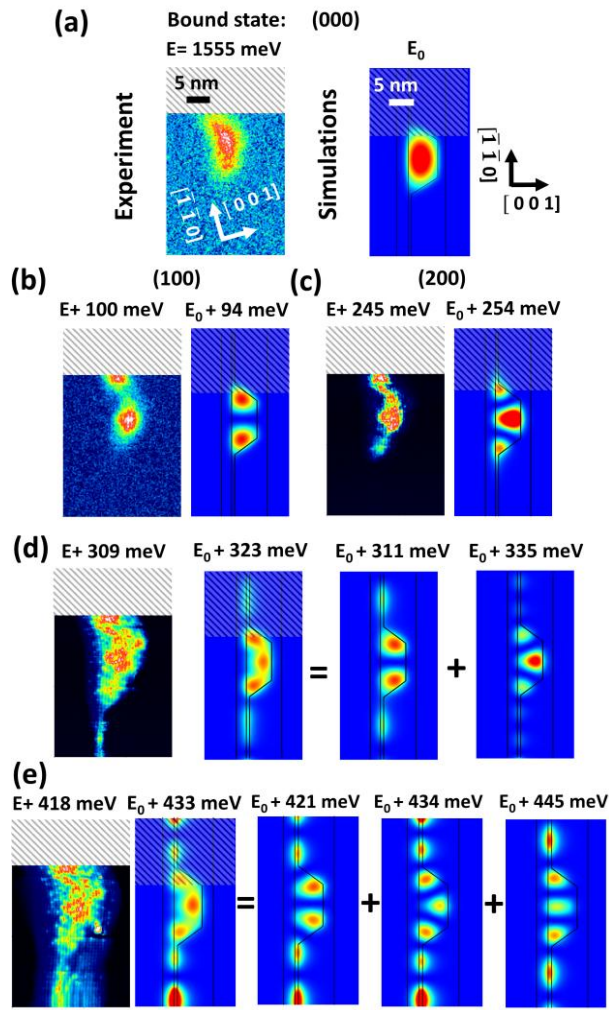

**Figure 3:** Real-space direct wave function imaging obtained from measured *dI/dV* maps of an individual QD in layer #7, along with simulations for a QD with a similar geometry. The upper shaded areas represent missing experimental data due to tip drift during the 12-hour acquisition time. To facilitate comparison, we marked the corresponding areas in the simulation plots. The simulations display the square modulus of the wave function on the plane separating the semiconductor from the vacuum, in analogy with the STM experiments. For the calculations we assumed a cleaved truncated pyramid QD geometry with an exposed QD width and height of 18 nm and 6.5 nm as seen from the STM tip, and a sub-surface depth of 9 nm. The measured energy level differences between successive bound states are in good agreement (up to 15 meV) with the modelled values (a-c). In the cases where two or more states are closer in energy than the energy level broadening in our STM setup (31 meV at 77 K - see Methods), their combined state is considered as shown in panels (d) and (e). Each panel has its own arbitrary colour scale.

## 2.4 Laser fabrication and assessment.

To obtain a more comprehensive evaluation of the quality and performance of our Si-based laser structure, we fabricated Fabry-Perot (FP) laser devices using the same epitaxial structure (a different piece from the same wafer) used in the STM/STS analysis. Figure 4a depicts the high-level process flow employed for the ridge-waveguide laser fabrication, whereas Figure 4b illustrates a schematic of the resulting device. The detailed fabrication processes for both narrow ridge-waveguide lasers, designed for single-transverse-mode operation, and broad area (BA) lasers, ideal for assessing material quality, are described in Methods. All lasers emit around 1.3 µm at room temperature. Figure 4c displays light-current (L-I) curves showing the temperature dependence of the threshold current for a BA laser with a cavity length of 2.8 mm and width of 50 µm. Its estimated CW characteristic temperature, $T_0$, is 51 K within the temperature range of 20-60 °C, and 32 K between 70-120 °C (inset of Figure 4c). The lasers generally demonstrate highly satisfactory CW performances, both at RT and at elevated temperatures around and above 80 °C. Notably, the as-cleaved III-V-on-Si broad area lasers with L = 2.8 mm exhibit a RT threshold current of 67 mA, which translates to a RT $J_{th}$ of 48 A cm$^{-2}$, corresponding to 6.85 A cm$^{-2}$ per QD layer. Moreover, they operate up to 125 °C and exhibit an output power per facet above 8 mW at 80 °C. Such high performance in BA lasers is a direct reflection of the high material quality in the active region, making it suitable for fabrication of more elaborate lasers, amplifiers, and other devices.

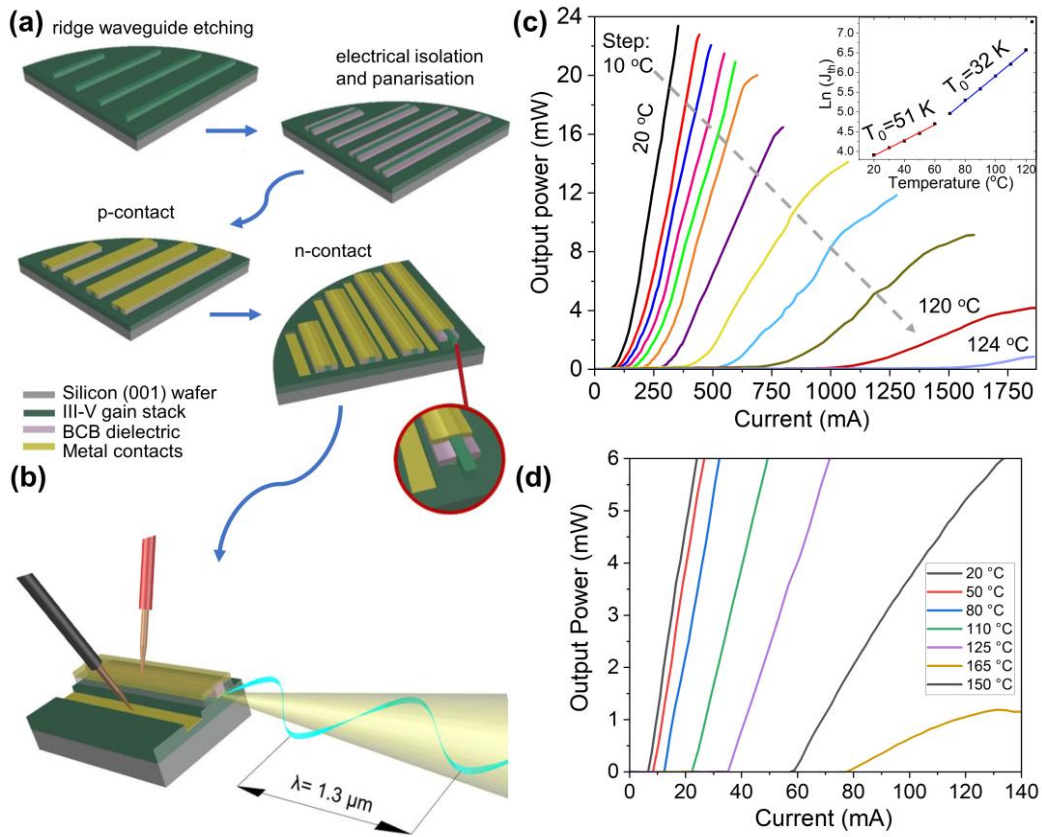

**Figure 4:** (a) High-level process flow describing the laser fabrication. (b) Schematic of the final fabricated laser. (c) L-I curves showing the temperature dependence of the threshold current of a broad area laser with a cavity length L = 2.8 mm and width w = 50 µm, under CW operation. Inset: The characteristic temperature $T_0$ extracted from this device was calculated as the inverse of the fitted slope. (d) CW L-I curves at various heat-sink temperatures for a ridge laser with one facet high-reflection (HR) coated and cavity length L = 1560 µm and width w = 2 µm. The emission wavelength is centred at $\lambda$ = 1.3 µm at room temperature. BCB—benzocyclobutene.

Another crucial requirement for lasers in SIP and next-generation datacom applications is the ability to maintain appreciable output power at elevated temperatures, particularly in FP single-transverse-mode operation—suitable for frequency combs—or in single-longitudinal (DFB) mode. Here, we processed and investigated FP single-transverse-mode ridge lasers, which fulfil both low-threshold and high-output-power requirements at 80 °C and even above. Our as-cleaved (before the application of facet coatings) narrow ridge lasers typically exhibited a CW output power per facet above 25 mW at RT and over 8 mW at 80 °C. Moreover, in a wire-bonded device with L = 2.2 mm, high-temperature CW ground state lasing was maintained up to 165°C, providing additional confirmation of an exceptional QD material quality and size uniformity in the active region. This performance compares favourably with the highest operating temperatures reported for Si-based lasers, which have reached 108 °C with undoped active media[11] and 119 °C using p-type modulation doping[42], a technique known to enhance temperature insensitivity but with the trade-off of increased threshold current. On native GaAs substrates, a maximum lasing temperature of 220 °C has been reported for lasers with both facets high-reflection (HR) coated[30]; however, these devices also incorporated p-type modulation doping, an approach we intentionally avoided here aiming for the lowest possible threshold currents at RT and 80 °C.

Moreover, a particularly high operating temperature of 150 °C was recently demonstrated on a Si(001) substrate[43], but this was achieved with a ridge width of 6 μm. While such a ridge width can generally support higher maximum temperatures, it significantly reduces laser efficiency and is incompatible with single-transverse-mode operation. For instance, the discrepancy in efficiency immediately becomes clear when comparing the threshold current at elevated temperatures between these lasers and ours. This example highlights that directly comparing individual laser performance characteristics from different literature sources can be misleading without considering the specific details of each device, especially the cavity dimensions. In particular, the maximum operating temperature, threshold current, and threshold current density are all strongly influenced by various design parameters, including cavity dimensions, doping profiles, and contacts, in a non-trivial manner. For this reason, we also focused on demonstrating that the quality of our active material, grown directly on Si, is comparable to, if not equal to, that of the same active region grown on a native III-V substrate. This is crucial because it implies that performance can be optimized to match that achieved on GaAs substrates, tailored to the needs of each specific application.

To this end, and in order to facilitate a direct comparison between our Si-based devices and their native GaAs-based counterparts, we processed lasers on both substrates as identically as possible. Specifically, we grew the same epitaxial structure as depicted in Fig. 1b on a GaAs substrate and then fabricated lasers using the same process on both samples simultaneously. Notably, the average lasing threshold of twelve III-V-on-Si lasers (w = 2 μm; L = 2200 μm), measured at 16.58 ± 1.78 mA, is comparable to virtually identical devices on the native GaAs substrate, which recorded 16.67 ± 2.23 mA. To our knowledge, this is the first instance where III-V-on-Si lasers demonstrate equivalent performance to III-V-on-III-V lasers that are identical in all respects—growth, fabrication, cavity dimensions, and facet coatings—except for the substrate. Similarly, other lasers from the same wafers also exhibited similar performance characteristics.

Notably, the application of an HR coating on one facet reduced the threshold current to 8 mA at RT and 17 mA at 80 °C for a 2 μm x 2200 μm cavity, and to 6 mA at RT and 12 mA at 80 °C for a 2 μm x 1560 μm cavity. The high-temperature output characteristics of these HR/as-cleaved devices improved accordingly, with the output power from the as-cleaved facet exceeding 35 mW at 80 °C. This output power surpasses that of previously reported Si-based lasers with similar cavity dimensions by more than an order of magnitude and is two to three orders of magnitude higher than other low-threshold current III-V-on-Si laser schemes, such as VCSELs or microring lasers[33,44]. Figure 4d illustrates L-I curves of the L = 1560 μm ridge laser at various heat-sink temperatures. We note that the device yield of our process is high, with the measurements showing repeatable performance across different dies. These results further substantiate the material quality as assessed using the X-STM/STS approach in the first part of the present study.

## 3. Discussion and Outlook

In this study, we have successfully grown optimised III-V-on-Si and III-V-on-III-V laser structures using MBE, rigorously analysed them using state-of-the-art experimental and numerical methods, and then fabricated lasers on both substrates, evaluated them, and compared them. Various important technological applications could benefit from highly-efficient III-V-on-Si integrated lasers, as the ones developed here. In this regard, an additional key feature of our approach is the direct growth on flat, unprocessed Si substrates, distinguishing it from other prominent III-V-on-Si growth techniques that rely on patterned substrates[45,46,47,48,49,50,51,52]. Eliminating the need for substrate patterning is potentially favourable in most applied scenarios as it improves scalability and mass production prospects. These factors, coupled with the demonstrated performance improvements, render our epitaxial approach highly versatile.

In the rapidly evolving landscape of information and communication technologies, there is a pressing need for high-density optical interconnects with low power consumption. The development of integrated lasers that operate efficiently at elevated temperatures on Si can play a pivotal role in meeting this demand by simplifying system design, reducing costs, and lowering power consumption. In this context, the low threshold currents and high output powers at and above 80 °C demonstrated in our laser devices are highly suitable for next-generation data communication and data storage applications. Moreover, the robust integration of III-V lasers with Si technology offers significant potential for sensing applications in both environmental and biological contexts.

From a materials perspective, the achievement of III-V-on-Si laser structures with near-optimal defect densities in the active region, rivalling those in state-of-the-art III-V-on-III-V lattice-matched growth, marks a significant milestone in III-V-on-Si direct heteroepitaxy and SiP. The high QD material quality and size uniformity, as revealed by our X-STM/STS experiments, directly translate to high-performance lasing. This breakthrough underscores the potential of III-V-on-Si direct growth methods for the seamless and scalable integration of photonics with Si-based CMOS electronics. Looking ahead, the integration of our growth technology with silicon waveguides and passive SiP devices appears promising and feasible through various methodologies. A particularly promising one, involves defining an etched window on a silicon-on-insulator wafer, as demonstrated in recent literature.[35,53,26]

Moreover, the potential to implement our growth developments on Si(001) substrates is of particular importance, as it would enable broader compatibility with CMOS processing and industrial production lines. Notably, recent research has demonstrated that by carefully adapting the nucleation (seed) layer Si(001) can be accommodated[26], an approach that seems particularly suitable for transferring the structure developed here onto Si(001). Other promising methods for growing lasers on Si(001) with significant progress in recent years include V-grooved structures[54], U-shaped patterned Si (001)[45], using a GaP interfacial layer[55], and direct nucleation of a GaAs film with specialized Si wafer preparation[56].

In conclusion, the advancements achieved in this study demonstrate that the performance limitations anticipated due to the GaAs/Si mismatch can be entirely overcome using meticulously optimised, yet well-established, technological processes. This opens up exciting new possibilities for III-V-on-Si monolithic integration, which could extend to various photonic and phononic devices[57,58], topological structures[59,60], and other material systems. Our approach paves the way for the development of highly efficient integrated photonic circuits on silicon, which can be leveraged to address crucial contemporary technological challenges. Furthermore, we believe that thanks to a higher tolerance to lattice mismatch, the diversity of materials enabled by direct growth on a flat Si substrate may offer a new platform for performing on-chip experiments for basic research. This could, in turn, be used to investigate more fundamental issues in photonic or quantum systems[61,62,63,64,65].

## 4. Methods

Materials

We grew the InAs/InGaAs/GaAs quantum dot-in-a-well (DWELL) laser stack using a solid-source molecular beam epitaxy (MBE) system, on an n-doped silicon (001) substrate with a 4° off-cut angle towards the [011] plane. To facilitate dislocation reduction and micro-crack elimination, we employed a combination of strategies in different layers of the structure. First, we grew a 6 nm-thick AlAs nucleation layer at a low temperature (390 °C) on an oxide-free Si surface achieved by annealing the substrate at 900 °C for 10 minutes inside the MBE chamber prior to growth. This nucleation layer, effectively suppresses three-dimensional growth, a major source of threading dislocations (TDs), while also providing a high-quality interface for subsequent III-V material growth.[66] Following this, we deposited a 30 nm GaAs layer at 330 °C and a rate of 0.1 monolayers (ML)/s, then increased the growth temperature to 510 °C for depositing a 270 nm thick GaAs layer, and finally annealed it to 590 °C for the remaining 600 nm of the GaAs buffer layer deposition. This three-step temperature growth of the GaAs buffer layer effectively suppresses the TD propagation through strain relaxation. However, a high density ($1 \times 10^9$ cm$^{-2}$) of TDs remains propagating towards the active region.

To further decrease the TD density and improve the quality of subsequent III-V layers, we grew strained-layer superlattices (SLS) to serve as dislocation filter layers (DFLs).[8,67] Each SLS consisted of five layers of n-type $In_{0.18}Ga_{0.82}As$/GaAs (10 nm/10 nm) grown at a low temperature of 500 °C. The SLS formation induces TDs bending, acting as a filter for TDs with a high degree of edge orientation, leading to a significant reduction in TD density. In our approach, a total of 5 SLS were incorporated, with successive SLS separated by 300 nm *in-situ*-annealed, high-temperature-grown (590 °C) GaAs spacer layers. For each of the five GaAs spacer layers, thermal annealing was performed by pausing the growth in the MBE reactor and increasing the substrate temperature to 660 °C for 6 minutes. This process aimed to increase the mobility of the defects, leading to their further annihilation before subsequent layer growth.

Following these material quality improvements, we also aimed to stack a higher number of QD layers in the active region, which was recently proposed theoretically for Si-based lasers,[68] having 7 QD layers instead of the 5 usually used. Specifically, the active region consists of a 7-layer dot-in-a-well (DWELL) structure sandwiched between 1400-nm n-type (bottom) and p-type (top) $Al_{0.4}Ga_{0.6}As$ cladding layers. In the DWELL, the QD is placed asymmetrically inside the quantum well. More specifically, each DWELL layer includes a 2 nm $In_{0.16}Ga_{0.84}As$ layer at its bottom, a 2.7 ML InAs QD layer grown on it, and then capped by a 5 nm $In_{0.16}Ga_{0.84}As$ layer to complete the 7-nm quantum well (QWell). The typical QD height, as deduced from the atomic rows of our STM images, is between 6-7 nm whereas their width varies between 17-22 nm. The barriers between successive DWELL layers were grown in two steps, and consist of a 5 nm low-temperature-grown (510 °C) GaAs and an *in situ* annealed 32.5 nm high-temperature-grown (590 °C) GaAs spacer layer, a combination we found to maximize the photoluminescence performance.

Additionally, four AlGaAs/GaAs superlattices of 1 nm/1 nm unit cell have been introduced above the dislocation filters, below and above the active region, and above the p-cladding before the growth of the p-contact layer (see Fig. 1b). This novel scheme of superlattices is shown to produce smoother surfaces and to eliminate some strain-induced micro-cracks that were seen during previous III-V-on-Si growths. Such micro-cracks substantially reduced the useful area of the sample and the high-performance laser yield. This improvement was confirmed by examining the top surface of the sample using optical and SEM microscopy and also by examining the topography of the intermediate layers using cross-sectional STM. To avoid excited state lasing, we meticulously optimised the QD growth conditions to achieve a high QD density of $5 \times 10^{10}$ cm$^{-2}$ with a homogenous size distribution. The QD layers have been grown at 510 °C, which gives the highest PL intensity at room temperature, with a full-width half maximum (FWHM) of 27 meV. The relatively low InAs deposition temperature also enabled larger QD sizes as well as density, and so it allowed to shift the QD RT emission wavelength from 1255 nm to 1300 nm.

To summarize, in comparison to previous demonstrations, our current epitaxial approach introduces the combination of multiple growth strategies resulting in an overall novel epitaxial structure (Fig. 1b), as well as the optimization of key growth steps individually. Specifically, it includes an optimization of the growth parameters for the DFL superlattices, the introduction of additional superlattices above the DFLs and above and below the active region, and the growth of 7 QD layers instead of 5 usually used in similar structures[8,68], all of which are shown to exhibit exceptionally low defect densities.

Device fabrication

(a) Broad area lasers.

We fabricated the 50-μm-wide stripes using photolithography followed by wet chemical etching to about 100 nm above the active region. This rather deep etching profile serves to achieve an improved carrier confinement. Ti/Au p-type contacts were deposited by thermal evaporation on the p$^+$-GaAs contact layer on the top of the structure (see Fig. 1b; Fig. 4a). Ni/GeAu/Ni/Au n-type contacts were evaporated on the n$^+$-doped GaAs buffer layer (see Fig. 1b) in an etched window beside the ridge (Fig. 4a). After lapping the silicon substrate to about 100 μm, we cleaved laser bars to various desired cavity lengths. Following the same technique developed by Chen et al.[8], we achieved very clean mirror-like facets. The lasers were then mounted as-cleaved (no coatings applied to the facets) onto copper heatsinks, with the epitaxial side facing up (epi-side-up) and gold-wire-bonded for electrical characterisation.

(b) Single-transverse-mode ridge-waveguide lasers.

Whilst broad area lasers are the most suitable for assessing the material quality of the structure, we have also fabricated shallow ridge lasers to investigate their single-transverse-mode performance. To optimize the design of the ridge geometry, we first obtained experimental refractive index parameters by growing specific layers on individual substrates and measuring them with spectroscopic ellipsometry.[69] In the fabrication process, the ridge width was set to 2 μm as follows. First, we defined 1.9 μm wide stripes of Ti/Au/Cr (200/2000/100 A°) using electron beam lithography (EBL). Subsequently, using a second EBL step aligned with the first one, 2.0 μm wide hydrogen silsesquioxane (HSQ) trenches of 300 nm thickness were placed on top of the metallic ones, to serve as the mask for the waveguide definition. By protecting the metallic layer under the HSQ mask, we ensure a good etching quality of the semiconductor, while having the advantage that the metallic layer was deposited prior to the mask, which ensures a better contact[70,71].

We etched the shallow ridge using a combination of two $Cl_2$-based inductively coupled plasma (ICP) processes, one optimised for GaAs and one for $Al_{0.4}Ga_{0.6}As$. We monitored the etching with an interferometric technique that allows to accurately stop the etching at the appropriate layer. We note that conveniently, during etching, some $SiO_2$ is sputtered on the sidewalls, coming from the base Si wafer in the ICP chamber on which the sample is placed. This thin layer acts as a passivation layer, preventing oxidation of the Al-containing layers of the ridge when in contact with the atmosphere. Subsequently, after the residual masks removal, we used thermally cured benzocyclobutene (BCB) resin for the planarization of the waveguides (see Fig. 4a). Ti/Au p-type contacts and Ni/GeAu/Ni/Au n-type contacts were deposited by thermal evaporation the same way as in the broad area laser process. Finally, we thinned the Si substrate to about 100 μm, cleaved the laser bars, mounted them epi-side-up on copper blocks, and gold-wire-bonded them for testing. The ridge lasers were generally tested with as-cleaved facets. On two of the ridge lasers we applied an HR coating on their rear facet.

Measurement techniques

a) Scanning tunnelling microscopy and spectroscopy.

In the X-STM/STS experiments, the sample was cleaved *in situ*, perpendicular to the [1-10] axis, within a preparation chamber under ultra-high-vacuum (UHV) conditions (P < $10^{-10}$ mbar), as depicted schematically in Fig. 1a. Subsequently, the sample was transferred to the pre-cooled STM head for imaging and spectroscopic measurements at liquid nitrogen temperature (T = 77K). Our X-STM/STS setup enables atomic-scale resolution topography imaging and high-energy-resolution (of the order of 10 meV) differential conductance *(dI/dV)* spectra. The resolution was adjusted accordingly to accommodate the level of detail required or to scan larger areas. We performed combinations of large-scale imaging over a wider area, and high-resolution analysis by focussing on smaller areas. The former allowed us to collectively observe, with the same tip conditions, the behaviour of multiple QDs in different layers (Fig. 2a-f), whereas the latter enabled detailed analysis of the electronic structure of individual QDs (Fig. 1d,e and Figure 3).

We utilized X-STM/STS to directly investigate a low-doped sample, as-grown for typical device processing, and then fabricated devices from the same sample, facilitating a direct correlation between STM measurements and device performance. This builds on recent advances in X-STM/STS,[40] which allowed to study the cross-section of embedded nanostructures (quantum dashes) within the active region of a laser structure. This advancement eliminates the need to grow dedicated, highly doped nanostructures for STM measurements. This approach is made possible because the carriers injected by the STM tip into the undoped active layers, escape into the p or n layers above and below, enabling the probing of the active layers' local density of states (DOS) without the need for heavy doping in the active region.

(b) Wave function mapping.

We obtained the electronic structure of the DWELL structure through an STS analysis as follows. Our STS analysis is based on the relation between the differential conductance and the local density of states of the sample at the position of the STM tip apex.[39] In our STS experiments, the differential conductance (*dI/dV*) is directly obtained by using a lock-in technique and modulating the sample voltage V by $V_{mod}^{rms} = 8\ mV$ at a frequency of 923 Hz with a time constant of 30 ms. Following the Tersoff and Hamann model[72,73] and taking into account the TIBB effect[74,75] inherent to spectroscopy on semiconducting surfaces, the measured *dI/dV* is proportional to the local density of states (LDOS(*E,x,y*)) of the cleaved buried nanostructures.[39,40,76] Squared wave function distribution in real-space can then be accessed by mapping both spatially- and energetically-resolved differential tunnelling conductance *dI/dV(V,x,y)* spectra, i.e., by accessing the spatial distribution of LDOS as a function of energy over a cross-sectional area.[39,40] This technique requires a continuous data acquisition for over 12 hours with nm-precision in positioning. To accommodate this requirement, we employed a system arrangement with ultra-high stability.

(c) Laser characterisation.

The laser device characteristics were measured under CW and pulsed (1% duty cycle and 1 μs pulse width) conditions. The output power was collected from a photodetector normal to the laser facet. The lasers characteristic temperature, $T_0$, was determined by taking the inverse of the fitted slope of the natural logarithm of the threshold current density as a function of temperature.

Energy resolution of our spectroscopic measurements.

The energy resolution of our STM setup is given by[39] $\delta E = \sqrt{(2.5\ eV_{mod})^2 + (3.3\ k_B T)^2}$. In our STS measurements, the differential conductance is directly obtained using a lock-in technique, with 8 mV modulation $V_{mod}^{rms}$ of V at 923 Hz. We then obtain $\delta E$ ≈ 20 meV at temperature of 4K and $\delta E$ ≈ 31 meV at 77K. More details on the STM energy resolution are provided by Fain et al.[39] and references therein.

Model implementation.

To support our experiments, we developed a finite-element numerical simulation package using the software COMSOL Multiphysics. First, the eight-band strain-dependent k.p model[77] was employed to obtain the conduction band edges, leveraging the precise information about the physical geometries of the QDs and QWells obtained from our atomic-resolution STM imaging (see Fig. 1). More specifically, using the tensor elements as input to an eight-band k·p model,[77,78] the local band edges were calculated, and the remaining parameters were obtained from Vurgaftman et al.[79] and Kumar et al.[78]. Subsequently, the one-band envelope function approximation with the strain modified effective mass in each material was used to calculate the electronic wave functions within the quantum dot-in-a-well regions and the associated confined energy levels. To ensure the reliability of our numerical simulations, we followed the technical implementation outlined by

Kumar et al.[78] and the methodology of Hoglund et al.,[80] making necessary adjustments to incorporate the specific material and geometry parameters obtained from our STM measurements.

As our main structure, we considered a truncated pyramid QD geometry with a height of 6.5 nm and a base length of 18 nm, which was cleaved in half. The dot height, length, and cleavage position were varied in subsequent simulations to thoroughly investigate their influences on the electronic structure. The QD was placed asymmetrically in a 9 nm QWell and a uniform, 0.75 nm wide, wetting layer (WL) was inserted in-between the first part of the QWell and the QD. We note that the WL is particularly important in determining the QD electronic structure, with a few previous reports considering it in detailed quantum mechanical calculations[81]. For this reason, we also performed additional runs using various WL widths to see how our wave function profiles are affected, with 0.75 nm being the optimal width to match the experimental data. The whole structure was inserted in a GaAs matrix of 200 nm x 200 nm, and a vacuum layer was added in front of the cleaved plane where the STM measurements are normally taken. More specifically, the axis of the pyramid is the [001] growth direction and the truncation plane corresponds to the (110) cleavage plane.

The QD/WL structure is surrounded by InGaAs (QWell), except on the side corresponding to the (110) surface which is exposed to vacuum (see Fig. 1a; Fig. 3). In our model we have disregarded the coupling between adjacent QDs, which we expect to be negligible for the vast majority of the dots in our sample.[41,82] Moreover, based on our experimental findings, we concluded that many-body effects could be disregarded as well. Specifically, this was supported by additional experiments we did using a 100 pA tunnelling current, corresponding to a 1 ns electron injection time, which is close to the electron lifetime in an un-cleaved QD. The boundary conditions were also chosen carefully to ensure a proper physical representation. The contact pair boundary condition was used between the interfaces of the heterostructures in the strain calculations. In the calculations for the eigenvalues, the Dirichlet boundary conditions were used on all outer surfaces of the geometry.

### Energy-levels superposition due to thermal broadening at 77K

As discussed in Figure 3d and e, when comparing our experimental wave function mapping with our simulated results, we considered the superposition of adjacent energy levels when their energetic distance was smaller than the thermal broadening at T = 77K. It is important to note that when plotting the simulated superposed wave function of multiple (>2) adjacent energy levels, as the ones of Figure 3e, we did not assume an equal weighting. The weighting was rather optimised such that the modelled wave function would look as similar as possible to the corresponding experimental one. This method assumes that the weighting in such superpositions depends on the energetic distance of each state from the selected energy at which the plot is shown. More specifically, in Figure 3e, the state at $E_0$ + 434 meV has a larger contribution due to its proximity to the selected energy ($E_0$ + 433 meV) at which the superposition is plotted. Since the corresponding experimental plots were also captured at specific energies, we expect that the experimental superpositions are also not weighted equally, and, thus, the comparison is fair.

## Acknowledgements

This work was supported by EU H2020 L3MATRIX Grant No. 688544, ANR FRONTAL Grant No. ANR-19-CE09-0017-02, and UK EPSRC EP/P006973/1, EP/T028475/1, and EP/X015300/1.

## Data Availability

The data that support the findings of our study are available upon request.

## Conflicts of interest

There are no conflicts to declare.